# Giant anisotropic magnetoresistance and nonvolatile memory in canted antiferromagnet $Sr_2IrO_4$

Haowen Wang[1], Chengliang Lu[1*], Jun Chen[2], Yong Liu[3], S. L. Yuan[1], Sang-Wook Cheong[4], Shuai Dong[2†], and Jun-Ming Liu[5,6]

[1] *School of Physics & Wuhan National High Magnetic Field Center, Huazhong University of Science and Technology, Wuhan 430074, China*

[2] *School of Physics, Southeast University, Nanjing 211189, China*

[3] *School of Physics and Technology, Wuhan University, Wuhan 430072, China*

[4] *Rutgers Center for Emergent Materials and Department of Physics and Astronomy, Rutgers University, Piscataway, New Jersey, 08854, USA.*

[5] *Laboratory of Solid State Microstructures and Innovation Center of Advanced Microstructures, Nanjing University, Nanjing 210093, China*

[6] *Institute for Advanced Materials, Hubei Normal University, Huangshi 435001, China*

---

* cllu@hust.edu.cn
† sdong@seu.edu.cn

# Abstract

Antiferromagnets have been generating intense interest in the spintronics community, owing to their intrinsic appealing properties like zero stray field and ultrafast spin dynamics. While the control of antiferromagnetic (AFM) orders has been realized by various means, applicably appreciated functionalities on the readout side of AFM-based devices are urgently desired. Here, we report the remarkably enhanced anisotropic magnetoresistance (AMR) as giant as ~ 160% in a simple resistor structure made of AFM $Sr_2IrO_4$ without auxiliary reference layer. The underlying mechanism for the giant AMR is an indispensable combination of atomic scale giant-MR-like effect and magnetocrystalline anisotropy energy, which was not accessed earlier. Furthermore, we demonstrate the bistable nonvolatile memory states that can be switched in-situ without the inconvenient heat-assisted procedure, and robustly preserved even at zero magnetic field, due to the modified interlayer coupling by 1% Ga-doping in $Sr_2IrO_4$. These findings represent a straightforward step toward the AFM spintronic devices.

**Introduction**

Antiferromagnets represent the overwhelming majority of magnetic materials in nature, and exhibit fascinating physical properties, such as diverse spin textures and topologically protected states of matter [1,2,3,4]. Although the antiferromagnetic (AFM) order serves as the second basic type of magnetic order, and has been discovered for nearly a century since the 1930s, it has so far been used as passive elements, such as pinning of the ferromagnetic (FM) layer in modern spintronics technology [5]. The main obstacle that keeps the AFM materials away from even more extensive applications is the great challenge in detection and manipulation of AFM orders because of their imperviousness to external magnetic field. Recently, this perception has been largely modified, and the emerging concept of AFM-spintronics has been garnering considerable interest [6,7,8,9].

The AFM-spintronics, where magneto-transport is governed by an antiferromagnet instead of a ferromagnet, opens perspectives for both fundamental research and device technology, since the AFM orders offer some unique and irreplaceable advantages compared to ferromagnets. One of the major appealing features is the promise of ultrafast writing speed (on picosecond time scale), enabled by the THz scale AFM resonance which is three orders of magnitude higher than the ferromagnets with GHz resonance frequency [10,11,12]. The control of AFM orders, which was thought to be precisely difficult, has been experimentally realized by various means, including electrical switching [1,13,14,15,16], optical excitation [11], and heat-assisted magnetic recording [17,18]. On the detection side, anisotropic magnetoresistance (AMR), which is the magneto-transport counterpart of magnetic anisotropy energy, has been generally and successfully utilized, in analogue to the traditional FM-based spintronics [7,18,19,20,21,22,23,24,25]. However, the readout signal, i.e. the AFM-based AMR (AFM-AMR), is commonly found to be small at a level of ~1% (or even smaller) [7,22]. This is incompatible with the scalability of the readout time and cell size in modern spintronics devices, and conceptual breakthrough on the readout side of AFM spintronics is required to obtain significantly larger signals.

Magnetocrystalline anisotropy energy (MAE) determined by spin-orbit coupling (SOC) has been utilized as a generic principle to realize the AFM-AMR in previous studies. For instance, when the AFM spin axis is aligned along different directions with respect to the crystal axes, various resistive states are expected because of the induced anisotropy of electronic structure [19,23,24]. Although calculations predicted large density of states (DOS) anisotropy in antiferromagnets with strong SOC, experimentally observed AMR ratio is unfortunately small overall [19, 20, 26]. The substantial gap between theories and experiments based on MAE solely indicates that additional ingredients should be involved to enhance the AFM-AMR. Regarding this, AFM junction may be an alternative, and indeed giant AMR exceeding 100% at low temperature ($T$) was reported in an AFM IrMn-based tunnel junction [25]. However, spintronics effects in AFM heterostructures are strongly sensitive to disorder and perfect epitaxy, which hindered their experimental realization [22, 27, 28].

Recently, a number of exotic phenomena, such as spin polarized current [29] and anomalous Hall effect [30, 31] which were assumed to only happen in ferromagnets, were revealed in bulk AFM materials. This provides an excellent opportunity to combine the typical spintronics phenomena, i.e. giant or tunneling magnetoresistance (GMR or TMR) with AFM-AMR without expense of structural complexity. Note that AMR is also a bulk effect. Such a scheme may harvest two advantages. First, significantly enhanced AFM-AMR is expected because the typical spintronics effects are usually remarkable. Second, the synergy of different types of magneto-transports may lead to unconventional AFM-based functionalities.

In the present work, we demonstrate the giant AMR reaching ~160% at low-$T$ in a simple resistor plate of spin-orbit coupled AFM $Sr_2IrO_4$ single crystal. Furthermore, by tiny Ga-substitution at Ir-site, nonvolatile resistive memory can be achieved, and the bistable states can be manipulated in-situ simply by varying the magnetic field (**H**) direction. Collaborative action of an atomic scale GMR-like effect and MAE is responsible for these super AFM-spintronics functionalities in this $J_{eff}=1/2$ antiferromagnet. In the first part of the article, we illustrate the physical descriptions of anisotropic magneto-transport in $Sr_2IrO_4$. In the second part, we present remarkably enhanced AMR and its reversal behavior. In the third part, we demonstrate the AFM nonvolatile memory effect in the tiny Ga-doped $Sr_2IrO_4$.

**Results**

**Fundamental aspects of the $J_{eff}=1/2$ antiferromagnet $Sr_2IrO_4$.** We first briefly recall some fundamentals of $Sr_2IrO_4$. The $J_{eff}=1/2$ magnetic moments, entangling both spin and orbital momenta due to the strong SOC (~ 0.5 eV) at Ir site, develop a canted AFM order with uniform in-plane canting angle $\phi \sim 12°$ below $T_N \sim 240$ K [32, 33, 34]. Neutron scattering demonstrated that the AFM state is in a twinned manner with equal domain populations [32, 33]. Because of the canting, net magnetic moments appear alternatively in each $IrO_2$ planes, and are coupled antiferromagnetically along the c-axis. Therefore, $Sr_2IrO_4$ is fully compensated at the ground state without macroscopic magnetization (**M**). A magnetic field **H** larger than the flop critical field $\mathbf{H}_{flop}$ drives the $J_{eff}=1/2$ moments flop transition into a weak FM phase in which the net moments of the $IrO_2$ layers are ferromagnetically aligned, accompanied concurrently with a sharp drop in resistance (R) at $\mathbf{H}_{flop}$ [23, 35, 36, 37, 38]. As revealed by microscopic investigations in $Sr_2IrO_4$, the basal-plane first nearest-neighboring exchange constant is about 60 meV [39], and the exchange interaction along the c-axis is only about 1 μeV [40]. Such large anisotropic exchange interaction in $Sr_2IrO_4$ makes the flop transition possible. The AFM to weak-FM transition in $Sr_2IrO_4$ resembles the GMR effect in traditional FM-based spin valves, and we call it the GMR-like effect. More details of the GMR-like effect can be found in the Supplementary information (see Supplementary Fig. 1). A similar effect was first revealed in AFM $La_2CuO_4$ [41], and then identified in double layered manganites [42]. This is an atomic scale effect and appears in a $J_{eff} = 1/2$ AFM state, which is unusual. While the previous studies of magnetotransport in $Sr_2IrO_4$ have illustrated some interesting features [19, 23, 24, 36, 37, 38], the AMR phenomenon related to the $J_{eff}=1/2$ AFM lattice remains elusive. One possible reason is related to the crystal quality, as revealed by Kim *et al.* [43]. For instance, evident in-plane magnetic anisotropy was identified very recently when the $Sr_2IrO_4$ crystal quality was improved [35, 43, 44].

**Abnormal anisotropic magneto-transport.** To check the physics highlighted above, we measured the MR = [$R(\mathbf{H}) / R(0) - 1$] data over a broad **H** (up to 5 T) and temperature T (35 K < T < 240 K) range, with the measuring geometry shown in the inset of Fig. 1. Here, electric current I is always applied along the [001] direction (the c-axis), and **H** is rotated within the basal plane with an angle $\Phi$ relative to the [100] direction. The magnetic easy axis is [100] with $\Phi = 0°$ and the in-plane hard axis is [110] with $\Phi = 45°$. Crystal quality is an important ingredient for observing evident anisotropy in $Sr_2IrO_4$ [43]. Therefore, X-ray diffraction (XRD) and energy dispersive spectroscopy (EDS) measurements were performed, which suggest that the samples are pure phase and high quality (see Spplementary Figs 2 and 3). It is seen that below $T_N$, the MR(**H**) curves at different $\Phi$ inter-cross at several **H**, as shown in Fig. 1a for $\Phi = 0°$ and 45° at three temperatures (For more MR data see Supplementary Fig. 4). A summarized phase diagram is shown in Fig. 1b which can be divided into three regions I, II, and III, respectively.

This phase diagram reflects the abnormal anisotropic magneto-transport behavior in $Sr_2IrO_4$. In fact, both $H_{c1}$ and $H_{c2}$ are intimately related to $H_{flop}$ where the MR drop emerges. This sharp MR drop is due to the atomic scale GMR-like effect associated with the $J_{eff}=1/2$ moments flop-transition. It is seen that the derived critical fields at $\Phi = 45°$ are approximately $\sqrt{2}$ times larger than those at $\Phi = 0°$ (see Supplementary Fig. 5), following a geometric relationship simply. This can be directly understood by considering the MAE. For instance, when $H$ is applied along the magnetic easy axis ($\Phi = 0°$), relatively low Zeeman energy gain is required to overcome the MAE, and thus to trigger the $J_{eff}=1/2$ moments flop transition and thereby the atomic-scale GMR-like effect. This is confirmed by the $M(H)$ data at various $T$, as shown in Fig. 2. The $J_{eff}=1/2$ moments flop transition is accompanied by a drastic enhancement in $M$ in the low field range for both $\Phi = 0°$ and $\Phi = 45°$, marking the conversion from the fully compensated state to the weak FM phase. In particular, it is true to see that the transition develops with a smaller critical field $H_{flop}$ at $\Phi = 0°$ (easy axis) than at $\Phi = 45°$ (hard axis), in consistent with the MR data. An advantage is a very small $H_{flop} \sim 0.2$ T, favorable for practical AFM devices with $Sr_2IrO_4$.

Furthermore, it is seen that $M$ at $\Phi=0°$ is overall larger than that at $\Phi=45°$ up to $H=5$ T (the maximum field used for $M(H)$ measurements in the present work). This can be seen up to $T_N$, demonstrating the unchanged magnetic easy axis after the flop transition [43, 45]. By plotting $\Delta M$ (=$M(\Phi=0°)$ - $M(\Phi=45°)$) as a function of $H$ (the insets of Fig. 2), the $M$-discrepancy can be seen more clearly, and the maximal $\Delta M$ is found at $H \sim H_{flop}$. However, at $H > H_{c2}$ shown in Fig. 1a, the relatively more conductive behavior is identified at $\Phi=45°$, i.e. $H$ // [110] which is the in-plane hard axis. Similar phenomenon was also identified in $Sr_2IrO_4$ thin films previously, and bandgap engineering due to rotation of the Ir moments was revealed to be responsible for this through first principles calculations [24]. For instance, the bandgap of $Sr_2IrO_4$ would be reduced when rotating Ir moments from the [100] direction (easy axis) to the [110] direction (hard axis) [19, 24]. This is physically anticipated considering the essential role of the strong SOC in build up the electronic structure in $Sr_2IrO_4$. In Fig. 1a, a clear slop-change can be seen at $H_{c1} < H < H_{c2}$ in the MR curves at $\Phi=45°$, after which the [110] direction is getting to be relatively more conductive than the [100] direction. This is indicative of the onset of the reduced bandgap dominating the transport. Complementary magneto-transport data measured under pulsed high magnetic field up to 50 T are shown in the Supplementary Fig. 6. It is seen that the unusual anisotropic magneto-transport due to bandgap engineering can persist up to very high field.

Regarding the MR bumps at **H**<**H**$_{flop}$, domain wall scattering may be qualitatively relevant based on three considerations. First, in region-I with **H**<**H**$_{flop}$, **H** is insufficient to trigger the flop transition, but may tilt the magnetic moments evidenced as gradual increase in **M**. However, the positive value of the finite MR in region-I excludes a possible origin arising from **H**-suppressed magnetic scattering effect, but indicates an enhancement of scattering effect in the system. Second, intensive domain wall motion has been demonstrated as **H**<**H**$_{flop}$ in SIO [43]. For instance, with increasing **H**//[100], the domain with $J_{eff}$=1/2 moments along [100] grows, but the domain with $J_{eff}$=1/2 moments along [010] shrinks. The domain repopulation involves complex domain wall motions reflected as deviations from linear behavior and hysteresis in the magnetization measurements, as shown in Supplementary Fig. 7. Meanwhile, evident MR-bumps with hysteresis is seen at **H**<**H**$_{flop}$, coinciding with the hysteresis effect in **M**(**H**). This close relationship among the MR(**H**), **M**(**H**), and domain wall motion leads us to propose a domain wall scattering scenario responsible for the observed MR-bumps. However, with **H**//[110], the domains remain populated as **H**<**H**$_{flop}$, since the field has no preference to the domains [43]. As shown in Fig. 1 and Supplementary Fig. 4, with **H**//[110], the MR-bumps are largely suppressed indeed, and instead continuous decrease can be seen in the MR curves. Simultaneously, the hysteresis in **M**(**H**) curves are also suppressed. These phenomena are consistent with the proposed domain wall scattering scenario. The remained weak MR-bumps may due to slight misalignment of **H** from the [110] direction, since any tilt of **H** leads to an imbalance in the domain population. Third, significant domain wall resistance has been theoretically revealed in SIO very recently by Lee et al. [37], further supporting the proposed domain wall scattering mechanism responsible for the MR-bumps in region-I. With increasing $T$, the MR bumps and its hysteresis are evidently suppressed at $T$>50 K, indicating that **H**-suppressed magnetic scattering gets to dominate the magnetotransport. Surely, further comprehensive investigations are desired to clarify the low-field MR in region-I.

**Giant anisotropic magnetoresistance and its reversal.** Having established the physical descriptions of the abnormal anisotropic magneto-transport in Sr$_2$IrO$_4$, we now are in a position to track the AMR effects in our simple AFM resistor. In Fig. 3, we plot $R$ as a function of $\Phi$ measured at $T$=35 K and $T$=90 K. Complementary AMR data can be seen in Supplementary Figs 8 and 9. Two features can be seen clearly, such as the **H**-driven AMR reversal and large $R$-variation at intermediate **H** region.

Fig. 3b and d show estimated AMR ratio (AMR=$R(\Phi)/R_{min}$-1) as a function of **H** obtained at $T$=35 K and 90 K, respectively. Here, $R_{min}$ means the minimal $R$ in the AMR traces. The AMR($\Phi$) curves exhibit pronounced peak around **H**~0.2 T. This is even remarkable at $T$=35 K, and the maximum AMR ratio is found to be as large as 160% which is the largest value compared to those reported so far in AFM materials. AMR phenomenon was studied using different experimental setup in $Sr_2IrO_4$ previously, and a detailed comparison with the present work can be seen in the Supplementary information (see Supplementary Fig. 10). As compared to the phase diagram shown in Fig. 1b, we can see that the striking AMR enhancement develops right within the region II (shadowed with olive). Therefore, the observed giant AMR effect can be ascribed to a combination of the atomic scale GMR-like effect and the MAE energy which dominate the magneto-transport together in this intermediate field region ($H_{c1}$<**H**<$H_{c2}$). In addition, inhomogeneities (i.e. domains and defects) within the sample resulting in regions with slightly different transition fields might also contribute a little to the AMR.

Here we note that the AMR symmetry in region II is more like twofold at both $T$=35 K and 90 K, although a fourfold rotation symmetry is in principle anticipated because of the square lattice of $Sr_2IrO_4$. As revealed by several recent theoretical works, the in-plane anisotropy in $Sr_2IrO_4$ can be accurately described by a magnetic-lattice coupling model [43, 45]. Importantly, it was found that the magnetic-lattice coupling effectively acts like a uniaxial anisotropy, once the magnetic moments are aligned along a certain direction, such as the $J_{eff}$=1/2 moments flopping. Therefore, it is physically reasonable to expect a twofold AMR symmetry at **H**~$H_{flop}$, such as the present case, according to the proposed magnetic-lattice coupling model. In addition, weak $R$-tips can still be seen in the basin of the AMR trace, indicating existence of retained biaxial anisotropy.

Having understood the drastic AMR enhancement in region II which is a key finding in the present work, we now turn to look at the AMR effect happening in Region I and III. Within region I, the magneto-transport is dominated by domain wall scattering as discussed above. For instance, clear hysteresis can be seen in the AMR traces within region-I, which may also duo to the domain wall motion akin to the MR-bumps. As revealed by Porras *et al.*, only the cases with **H**//[110] are not expected to cause domain repopulation [43]. Therefore, the AMR hysteresis appears at both easy- and hard-axes. In Fig. 3a, the induced AMR symmetry at **H**=0.1 T is twofold with peaks at $\Phi \sim 45°$ and $\sim 225°$. First, the application of **H**=0.1 T is insufficient to trigger the flop transition at $T$=35 K. Second, the domain wall motion becomes more impetuous when applying **H** along the magnetic easy axes (i.e. at $\Phi=0°$) than along the hard axes (i.e. at $\Phi \sim 45°$). Therefore, rotating **H** =0.1 T from hard- to easy-axis would significantly promote the domain wall motion, giving rise to the increase in $R$. If taking a closer look at the AMR trace, step-like anomalies with hysteresis can be identified (indicated by arrows), marking the booming of domain wall motion. Such domain wall motion process will be continued as further rotating **H** from $\Phi=0°$ to 45°, since only the cases of **H** along the hard axes do not change the domain population. This gives rise to the ongoing increase in $R$ till $\Phi=45°$. From $\Phi=45°$ to 135°, the **H** direction is partly reversed, and thus the domain walls are gradually removed. As a consequence, $R$ decreases with $\Phi$ from 45° to 135°. This process will be repeated when subsequently rotating **H**=0.1 T from $\Phi=135°$ to 315°, leading to the twofold AMR symmetry. In addition, such tunneling AMR is not observable at $T$=90 K (Fig. 3c), since the thermal fluctuation is sufficiently high to overcome the energy barrier at domain walls for electron transport. This is consistent with the phase diagram shown in Fig. 1b.

Regarding the fourfold AMR symmetry at **H**=3 T, it can be ascribed to the bandgap engineering akin to the analogous phenomenon in $Sr_2IrO_4$ thin films [19, 24]. Because **H**=3 T is much larger than **H**$_{flop}$, the weak FM phase is well stabilized, and its net magnetic moment and the corresponding canted AFM orders can (mostly) follow **H** rotating within the basal plane of $Sr_2IrO_4$. For instance, the Ir moments can factually reach both the [100] direction and the [110] direction when rotating **H** within the *ab*-plane. In this sense, the anisotropic magneto-transport dominated by bandgap engineering emerges, for example it is relatively more conductive as Ir moments pointing along the [110] direction (with smaller bandgap) compared with the [100] direction (with larger bandgap) [19, 24]. As a consequence, a fourfold AMR symmetry with minima appearing right at the in-plane hard axes is obtained. More details of giant AMR phenomenon and its reversal can be found in Supplementary Fig. 11.

**Nonvolatile memory in tiny Ga-doped $Sr_2IrO_4$.** The above data have shown a highly efficient scheme, which is the combination of various types of magneto-transports dominated by either spin polarized current or MAE energy, to enhance the AFM-AMR effect in $J_{eff}$=1/2 $Sr_2IrO_4$. Along this line, a natural proposal is if the various magneto-transport behaviors can be preserved after removing **H**. By this, we should be able to observe the interesting AFM-based spintronics functionalities at **H**=0 T, i.e. the AFM memory effect. A key step is to retain the flopped magnetic state from recovery after the **H**-writing. It is found that the non-doped $Sr_2IrO_4$ is not proper to realize this purpose, since the pristine state can be recovered after different **H** sweeping (Fig. 1a). We then dope $Sr_2IrO_4$ with 1% Ga at Ir-site, which was found to be efficient in triggering the flop transition without breaking the $J_{eff}$=1/2 state akin to the effect of applying external **H** [46]. As shown in Fig. 4a, it is true to see evident remnant magnetization $M_r$ in the **M**(**H**) curves taken at $T$=10 K. Importantly, $M_r$ at $\Phi$=45° is found to be much larger than that at $\Phi$=0°. The reason includes that first the larger $M_r$ means the stronger suppression of magnetic scattering. Second, more Ir moments remaining at the [110] direction would lead to relatively smaller bandgap in the resistor. Both effects are all pointing to a higher conduction along the [110] direction at **H**=0 T, as compared with the [100] direction in the tiny Ga-doped $Sr_2IrO_4$.

This is confirmed by further transport characterizations. We first zero field cooled the resistor to $T$=10 K, and then measured the MR by performing successive **H**-cycles either at $\Phi$=0° or $\Phi$=45°. As shown in Fig. 4b, after the first **H**-cycle, two reversible resistive states can be obtained by further **H** writing procedures, demonstrating an AFM memory effect. The two memory states can be switched by in-situ **H**-writing without heat-assisted procedure, and the resistive ratio of the two states is as large as ~4.5%. This is a major point deserved for high appreciation and it is different from the operation utilized in AFM MnTe [17] and FeRh [18]. The drift effect seen in the resistance over time (Fig. 4c) may due to inhomogeneities (such as domains) within the sample. Full $R$(**H**) data can be seen in the Supplementary Fig. 12. The memory effect is nonvolatile and fully reproducible in the successive write-read cycles, as shown in Fig. 4c. The same protocol was repeated several times with **H** applied initially either at $\Phi$=45° or $\Phi$=0°, and the same results were obtained. In our devices, the memory states are stable against a field up to ~1 kOe. Although this is smaller than that in compensated AFM materials showing very high rigidity, it is robust against electromagnetic disturbances, and the relatively small switching field would facilitate the operation in devices.

Having demonstrated the AFM-based nonvolatile memory in the tiny Ga-doped $Sr_2IrO_4$, we then go to some details of the physical properties of the resistor. First, both the high and low resistances ($R_H$ and $R_L$) are smaller than the pristine state. This can be understood directly by the different $\mathbf{M}_r$ for these cases. For instance, the initial state after zero field cooling has zero net magnetization [46]. Second, $\mathbf{M}$ at $\varPhi=45°$ is larger than at $\varPhi=0°$, different from the case in the non-doped $Sr_2IrO_4$. This should be ascribed to the modification of interlayer coupling by Ga-doping. As revealed by recent calculations, the interlayer coupling can effectively change the anisotropic magnetization in $Sr_2IrO_4$ [43, 45, 47]. If considering bare effect of interlayer coupling in $Sr_2IrO_4$, the [100] direction is easier than the [110] direction as AFM interaction dominates. However, this situation is reversed, i.e. the [100] direction behaves even harder than the [110] direction, when the interlayer coupling becomes FM-like (which can be realized by the flop transition). The significant role of interlayer coupling in determining anisotropic magnetism looks to be general in iridates [48]. As shown in Fig. 4a, the clear $\mathbf{M}(\mathbf{H})$ hysteresis evidences the appearance of FM coupling of the $IrO_2$ layers, in analogous to the Rh-doping effect [49]. Therefore, it is physically reasonable to expect different anisotropic magnetization in the tiny Ga-doped $Sr_2IrO_4$ compared to the non-doped one.

**Discussion**

We have demonstrated giant AMR and nonvolatile memory in simple AFM resistors made of $J_{eff}=1/2$ antiferromagnets without auxiliary reference layer. Note that coupling to FM layer would lose the unique merits of antiferromagnets in AFM spintronics devices. The AMR ratio is found to reach ~160% at $T=35$ K in the present work, which is much larger than previously available values [17, 18, 19, 20, 23, 37]. It is noted that very large tunneling AMR exceeding 100% at $T=4.2$ K was identified in AFM IrMn-based tunnel junctions, while additional FM NiFe layer was used for inducing exchange spring effect [25], different from the present case. Regarding the AFM-based nonvolatile memory, it has been rarely reported before. In the previous works, to obtained different memory states, the sample has to be heated above $T_N$, and then cooled below $T_N$ with magnetic field [17, 18]. Here we show an in-situ manipulation of the bistable states, which may have advantages in practical devices.

Our work has shown an efficient mean to achieve significantly large AMR and nonvolatile memory in the $J_{eff}=1/2$ antiferromagnets. The crucial part of our scheme is to combine various types of magneto-transport behaviors, i.e. the atomic scale GMR-like effect due to spin polarized current, and crystal AMR due to the MAE energy, in a single system. Importantly, the scheme proposed in the present work is applicable to a broad class of AFM materials. As a fundamental piece of the two major findings, the spin polarized current was theoretically predicted to generally exist in most antiferromagnets except those with simple collinear spin orders very recently [29]. In addition, the scheme could also be extended to some recently discovered AFM alloys where exotic anomalous Hall effect was identified, and distinctive AFM spintronic functionalities would be expected [30, 31, 50].

In summary, significantly enhanced AMR reaching ~160% at $T$=35 K is revealed in a simple resistor made of AFM $Sr_2IrO_4$ hosting $J_{eff}$=1/2 moment due to strong spin-orbit coupling, owing to a combination of an atomic scale GMR-like effect and the magnetocrystalline anisotropy. The enhanced AMR can be observed at higher-$T$, although its magnitude decreases gradually with increasing $T$. Further experiments demonstrated that the plural magneto-transport behaviors can be preserved by only 1% Ga-doping in $Sr_2IrO_4$ without the assist of external magnetic field. Importantly, in the tiny doped resistor, nonvolatile AFM memory effect is identified, and the bistable memory states can be operated in-situ. The modification of interlayer coupling by Ga-doping is essential for realizing the nonvolatile memory. Our work has illustrated a very efficient scheme to significantly improve the spintronics functionalities in antiferromagnets.

**Methods**

**Sample preparation.** The $Sr_2IrO_4$ single crystals ($2\times 1\times 0.5$ mm$^3$) were synthesized from off-stoichiometric quantities of $SrCl_2$, $Sr_2CO_3$, and $IrO_2$ using the self-flux techniques. The thoroughly mixed powders were placed in a platinum crucible covered with a lid, and then melted at 1250 °C in a programmable furnace in air. After this, the crucible was cooled to 800 °C at a rate of 6 °C per hour and then furnace-cooled to room temperature. The molar ratio of $SrCl_2$:$IrO_2$ was set at ~8:1, which was found to be critical for obtaining high quality crystals. For a comparison, 1% Ga-doped $Sr_2IrO_4$ single crystals were also synthesized through the same method. The stoichiometry of the resulting crystals was confirmed using energy dispersive spectroscopy measurements. The pure phase of the crystals was examined by performing room-temperature powder X-ray diffraction measurements on thoroughly crashed crystals. The crystals were also checked using Rigaku XtaLAB miniTM diffractometer at room temperature.

**Magnetization and electric transport measurements.** Electric transport measurements for the crystals were carried out using a four-probe method in a Quantum Design (QD) physical property measurement system equipped with a rotator module. Silver paste was used to make the electrodes. In order to have homogeneous electric current $I$ flowing through the sample, the electrodes of current source were made as large as possible (>1 mm$^2$). With regard to the magneto-transport measurements, exciting current $I$ was applied along the [001] direction, and the magnetic field $H$ was always applied within the basal plane of the crystals. Therefore, the AMR refers to the ($I$, **H**) behaviors. Magnetization as function of $T$ and **H** were measured using a QD superconducting quantum interference device. All the **M**(**H**) curves were measured after the zero-field cooled (ZFC) sequence. For the high field transport measurements, the same device structure was used, and the pulsed magnetic field up to ~50 T with a duration time of ~50 ms was generated by using a nondestructive long-pulse magnet energized by a 1 MJ capacitor bank.

**Data availability.**

The data supporting these findings are available from the corresponding author on request.

**Additional Information**

**Supplementary Information** accompanies this paper at doi:

**Competing interests:** The authors declare no competing interests.

**Acknowledgements:** We thank Prof. J. Liu for fruitful discussions and Prof. W. Li for single crystal characterizations. This work is supported by the National Nature Science Foundation of China (Grant Nos. 11774106, 11674055, 51431006, and 51571152), the National Key Research Projects of China (Grant No. 2016YFA0300101). S.W.C. is funded in part by the Gordon and Betty Moore Foundation's EPiQS Initiative through Grant GBMF4413 to the Rutgers Center for Emergent Materials, and also by the visiting distinguished professorship of Nanjing University sponsored by the State Administration of Foreign Experts Affairs of China.

**Author contributions:** C.L.L. conceived the project. H.W.W. prepared the samples and performed the physical property measurements. Y.L. participated the measurements. J.C., S.L.Y., and S.W.C. contributed to the analysis of the results. C.L.L., J.M.L., and S.D. wrote the manuscript.

**Figure captions**

**Figure 1. Magnetotransport in Sr$_2$IrO$_4$ single crystal.** (a) Measured magnetoresistance as a function of **H** applied at $\Phi=0^\circ$ and $\Phi=45^\circ$ at $T=35$ K, 50 K, and 90 K. The intercross points indicated by blue ($\mathbf{H}_{c1}$) and orange ($\mathbf{H}_{c2}$) dashed lines are used for building up the phase diagram. The **H**-sweeping directions are indicated by black arrows. For a better version, the curves have been shifted vertically. (b) Phase diagram of the anisotropic magneto-transport in Sr$_2$IrO$_4$ single crystal. The inset: the measurement geometry and two major crystalline directions of the crystal.

**Figure 2. Magnetic properties of Sr$_2$IrO$_4$.** Magnetization as a function of **H** applied at $\Phi=0^\circ$ and $\Phi=45^\circ$ at (a) $T=10$ K, (b) 90 K, and (c) $T=220$ K. The inset shows the difference of magnetization measure at two directions.

**Figure 3. Giant antiferromagnetic AMR and its reversal.** Anisotropic magnetoresistance measured with various **H**=0.1 T, 0.25 T, and 3 T at (a) $T=35$ K, and (c) $T=90$ K. The measurements were performed by sweeping **H** from -45° to 315° (red curves), and then back to -45° (black curves). Derived AMR ratio as a function of **H** obtained at (b) $T=35$ K and (d) $T=90$ K. The orange lines are used to highlight the remarkable AMR enhancement. Regions I, II, and III are labeled according to the phase diagram shown in figure 1.

**Figure 4. Nonvolatile memory and the in-situ control.** (a) Measured **M**(**H**) curves at $\Phi=0^\circ$ and $\Phi=45^\circ$ at $T=10$ K for 1%-Ga doped Sr$_2$IrO$_4$, in which different remnant magnetization can be seen. (b) $R(H)$ curves measured by successive **H** cycling along different directions at $T=10$ K. For instance, $\Phi=0^\circ\_1$ means the first **H**-cycle at $\Phi=0^\circ$. The field sweeping direction is indicated by arrows. The blue and red dots represent the high ($R_H$) and low ($R_L$) resistance state which can be switched by **H**-cycling. (c) Stable and reproducible memory states in-situ controlled by **H** at $T=10$ K. The corresponding net moment arrangements are schematically shown in the inset. In comparison with the case at $\Phi=45^\circ$, smaller net magnetization exists at $\Phi=0^\circ$ as **H**→0.

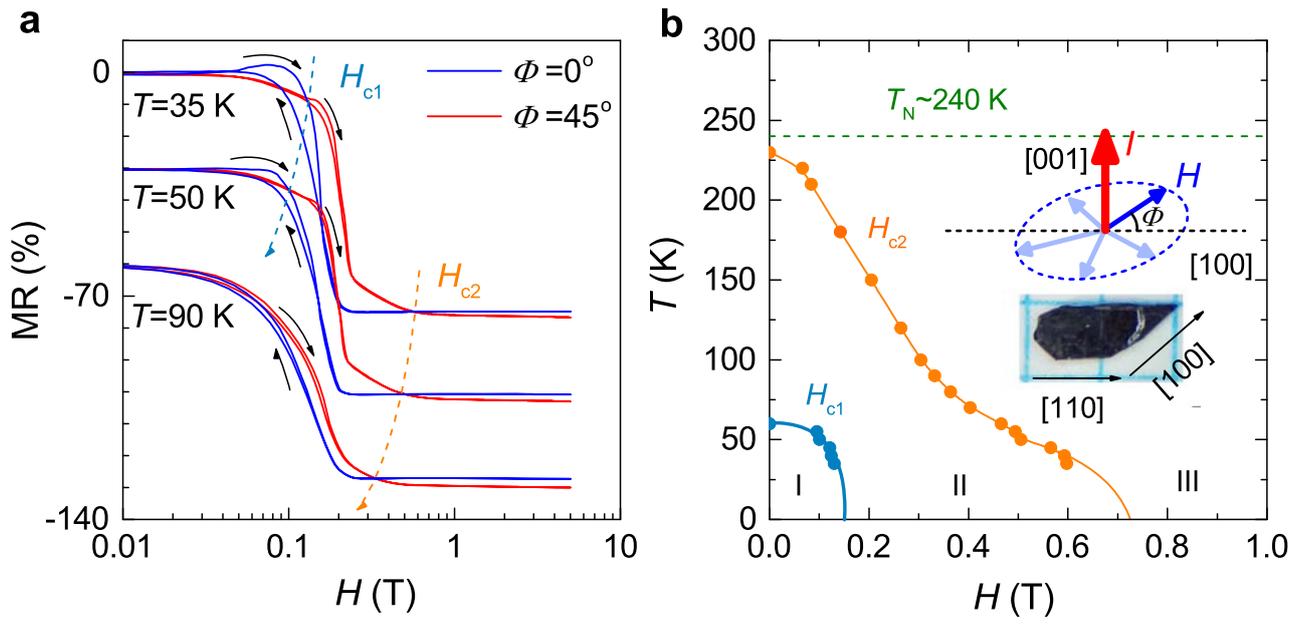

Figure 1

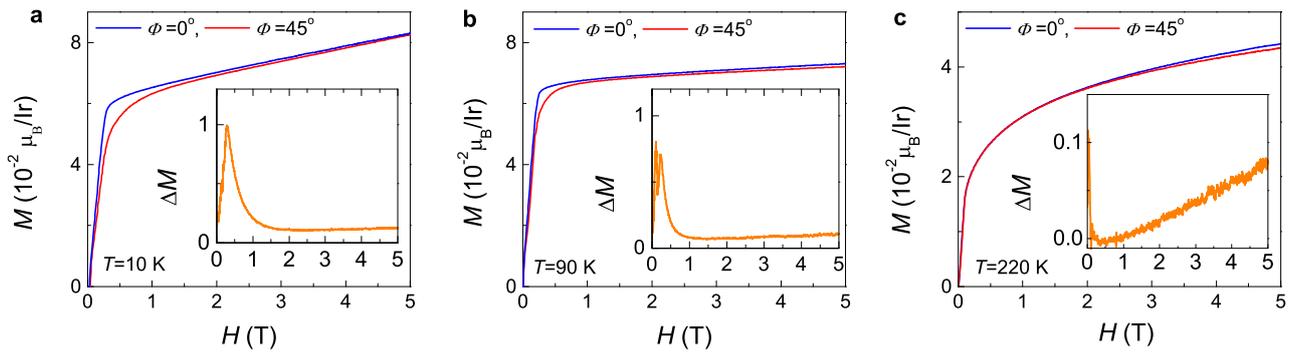

Figure 2

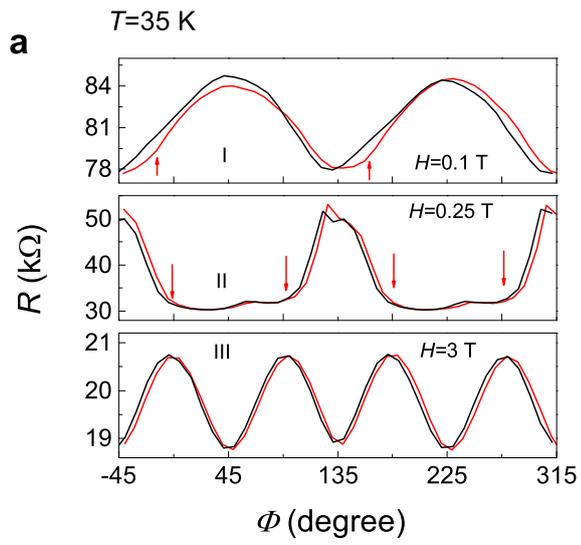
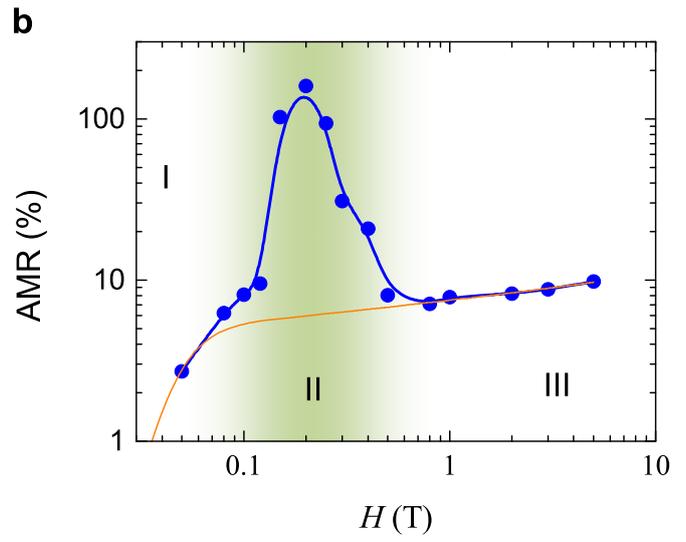
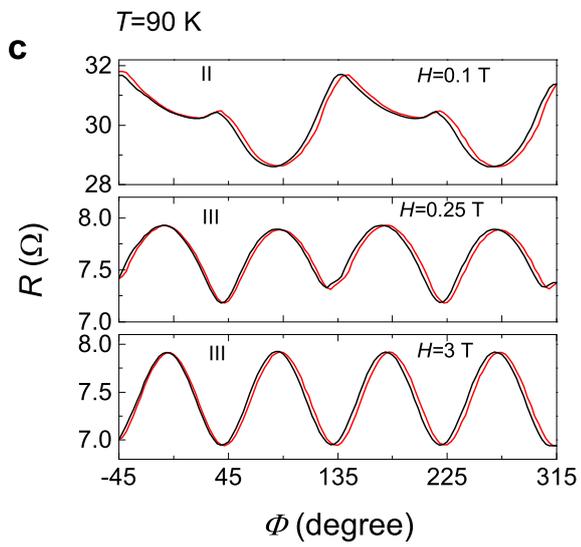
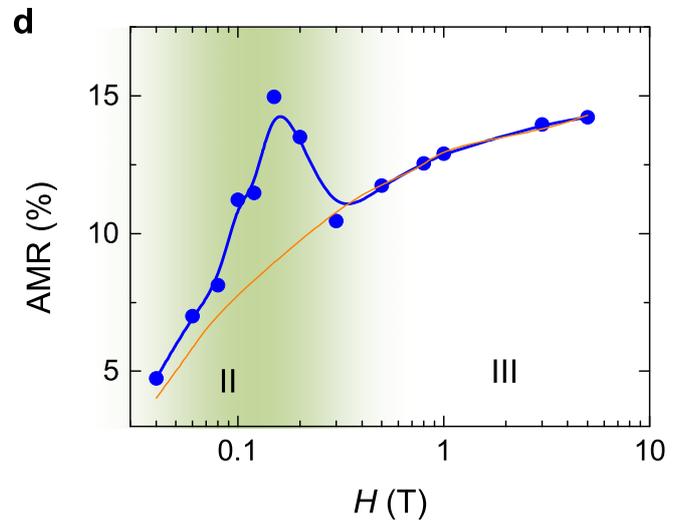

Figure 3

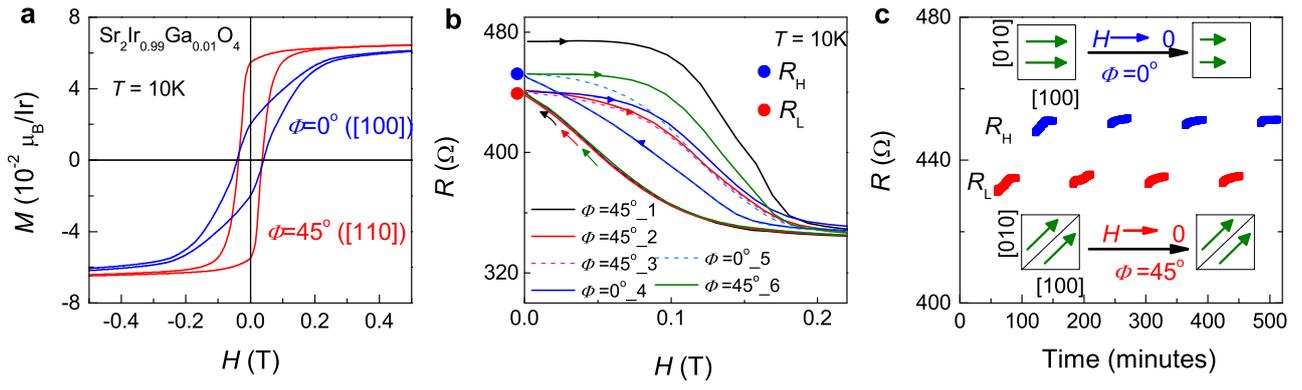

Figure 4